\documentclass[prl,twocolumn]{revtex4-1}
\usepackage{bm,graphicx,graphics,amsmath,amssymb,bm,epsfig,color}
\usepackage{euscript,tabularx}
\usepackage{textcomp}
\usepackage{gensymb}

\graphicspath{{./figs/}} 
\def\<{\left\langle}
\def\>{\right\rangle}
\def\unit#1{{\hat{\bm#1}}}
\def\op#1{{\widehat{\bm#1}}}

\begin{document}

 \title{Measurement of the dynamical dipolar coupling in a pair of
   magnetic nano-disks\\ using a Ferromagnetic Resonance Force
   Microscope}

\author{B. Pigeau}
\email{Corresponding author: benjamin.pigeau@cea.fr}
\author{C. Hahn}
\author{G. de Loubens}
\author{V. V. Naletov}
\altaffiliation{Physics Department, Kazan Federal University,
  Kazan 420008, Russian Federation}
\author{O. Klein}
\affiliation{Service de Physique de l'{\'E}tat Condens{\'e} (CNRS URA 2464), CEA Saclay, 91191 Gif-sur-Yvette, France}
\author{K. Mitsuzuka}
\author{D. Lacour}
\author{M. Hehn}
\author{S. Andrieu}
\author{F. Montaigne}
\affiliation{Institut Jean Lamour, UMR CNRS 7198, Universit{\'e} H.Poincar{\'e}, 54506 Nancy, France}

\date{\today}

\begin{abstract}
  We perform an extensive experimental spectroscopic study of the
  collective spin-wave dynamics occurring in a pair of magnetic
  nano-disks coupled by the magneto-dipolar interaction. For this, we
  take advantage of the stray field gradient produced by the magnetic
  tip of a ferromagnetic resonance force microscope (f-MRFM) to
  continuously tune and detune the relative resonance frequencies
  between two adjacent nano-objects. This reveals the anti-crossing
  and hybridization of the spin-wave modes in the pair of disks.  At
  the exact tuning, the measured frequency splitting between the
  binding and anti-binding modes precisely corresponds to the strength
  of the dynamical dipolar coupling $\Omega$. This accurate f-MRFM
  determination of $\Omega$ is measured as a function of the
  separation between the nano-disks. It agrees quantitatively with
  calculations of the expected dynamical magneto-dipolar interaction
  in our sample.
\end{abstract}

\maketitle

Studies of the collective dynamics in magnetic nano-objects coupled by
the dipolar interaction has recently attracted a lot of attention
\cite{chou06,loubens07a,gubbiotti08,awad10a,jung10,vogel10,sugimoto11,ulrichs11}
due to its potential for creating novel properties and functionalities
for the information technology. It affects the writing time of closely
packed storage media \cite{kruglyak10}, the synchronization of spin
transfer nano-oscillators \cite{belanovsky12}, and more broadly the
field of magnonics \cite{kruglyak10a}, which aims at using spin-waves
(SW) for information process \cite{karenowska12}. Despite the generic
nature of the dynamic magneto-dipolar interaction, which is present in
all ferromagnetic resonance phenomena, its direct measurement has been
elusive because it is difficult to reach a regime where this coupling
is dominant. It requires that the strength of the dynamical coupling
$\Omega$ exceeds both the deviation range of eigen-frequencies between
coupled objects and the resonance linewidth. Large $\Omega$ are
usually obtained by fabricating nano-objects having large
magnetization and placed nearby. But the constraint of fabricating two
nano-objects, whose SW modes both resonate within $\Omega$, is
difficult to meet. For long wavelengths, the SW eigen-frequency is
indeed very sensitive to imperfections in the confinement geometry,
inherent to uncertainties of the nano-fabrication process. Moreover, a
direct determination of the coupling strength between any two systems,
as for instance a superconducting qubit and electronic spins
\cite{kubo11}, requires the ability to tune and detune them at least
on the $\Omega$-range. So far, the absence of a knob to do so with the
individual frequencies of nearby magnetic objects has prevented a
reliable measurement of the dynamical dipolar coupling.

In this paper we shall demonstrate that ferromagnetic resonance force
microscopy (f-MRFM) allows this quantitative measurement of
$\Omega$. We shall rely on the field gradient of the magnetic tip as a
mean to fully tune and detune the resonance frequencies of two
nano-disks by continuously moving the tip laterally above the pair of
disks. One can find a position where the stray field of the tip
exactly compensates the deviation of internal field due to the
patterning process. At this position, the splitting between the
eigen-frequencies of the binding and anti-binding modes is exactly
equal to the dynamical dipolar coupling $\Omega$. By studying $\Omega$
as a function of the separation between the nano-disks, we shall
demonstrate that f-MRFM provides a reliable mean to measure the
strength of the dynamical coupling. It provides also a reliable mean
to measure mode hybridization and mode linewidth.

\begin{figure}
\includegraphics[width=8.5cm]{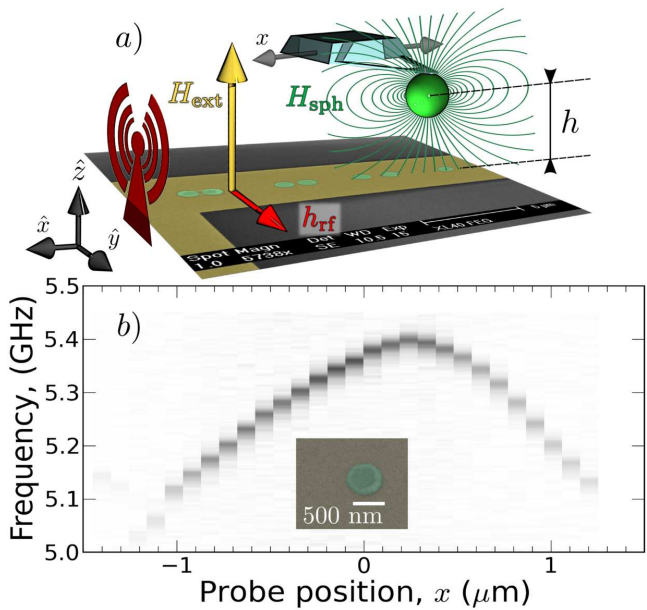}
\caption{a) Schematic of the f-MRFM setup: an Fe sphere glued at the
  apex of a soft cantilever is scanned laterally above different pairs
  of Fe-V disks excited by a microwave field. b) Density plot of the
  f-MRFM signal as a function of the displacement $x$ of the sphere
  above an isolated disk. The inset is a SEM image of the $2R=600$~nm
  Fe-V disk (green) placed below the microwave antenna (gold).}
\label{setup}
\end{figure}

The magnetic material used for this study is a $t=26.7$~nm thick Fe-V
(10\% V) film grown by molecular beam epitaxy on MgO(001)
\cite{bonell09,mitsuzuka12}. This is a ferromagnetic alloy with a very
high magnetization, $4 \pi M_s = 1.7 \times 10^4$~G, and a very low
magnetic Gilbert damping, $\alpha = 2 \times 10^{-3}$. The film is
patterned into disks by e-beam lithography and ion milling
techniques. The geometrical pattern (image in FIG.1a) consists in
three pairs of nearby disks having the same nominal diameter $2R=
600$~nm but different edge to edge separation: $s=200$~nm, 400~nm and
800~nm. Each set is separated by 3~$\mu$m in order to avoid cross
coupling. An isolated disk of identical diameter is also patterned for
reference purpose. The sample is then placed in the room temperature
bore of an axial superconducting magnet. The disks are perpendicularly
magnetized ($z$-axis) by an external field of 1.72 Tesla
\cite{angle}. This field is sufficient to saturate all the disks. A
linearly polarized ($y$-axis) microwave field $h_\text{rf}$ is
produced by a broadband Au strip-line antenna of width 5 $\mu$m
deposited on top of a 50~nm thick Si$_3$O$_2$ isolating layer, above
the magnetic disks.
The f-MRFM experiment consists in detecting the mechanical motion
produced by the magnetization dynamics in the Fe-V nano-disks of a
Biolever cantilever with an Fe nano-sphere of diameter 700~nm glued at
its apex (see FIG.1a) \cite{klein08}. We will consider in the
following that the stray field of the tip $\bm{H}_\text{sph}$ reduces
to the dipolar field created by a punctual magnetic moment
$m_\text{sph} = 3\times 10^{-10}$~emu placed at the center of the
sphere. The role of the magnetic tip in f-MRFM is to create a field
gradient tensor $\op G = \nabla \bm{H}_\text{sph}$ on the sample in
order to spatially code the resonance frequency and to provide a local
detection \cite{lee10}.

These two features are illustrated by FIG.1b, which shows the
dependence of the f-MRFM signal measured above the isolated disk as a
function of the position of the tip on the $x$-axis. It displays the
behavior of the lowest energy SW mode, where all spins are precessing
in phase at the Larmor frequency around the unit vector $\unit z$. The
cantilever is scanned at constant height $h$ above the sample
surface. The position $x=0$ corresponds to placing the probe on the
axis of the disk.

We first concentrate on the variation of the FMR resonance frequency
as a function of the $x$-position of the sphere. It displays a bell
curve, whose shape is due to the additional bias field produced by the
tip
\begin{equation} \label{eq:omega} \omega(x) = \omega_\text{FMR} +
  \gamma \{ {H}_\text{sph,z} (x)\} \,\, ,
\end{equation}
where the first term is the resonance frequency in the absence of the
sphere and the second term is the gyromagnetic ratio $\gamma$ times
the spatial average of the $z$-component of the stray field of the
sphere over the disk volume. The curly bracket indicates that this
average should be weighted by the spatial profile of the lowest energy
SW mode \cite{Bessel}. The maximum shift of frequency occurs close to
$x=0$, where the additional field from the f-MRFM sphere is maximal
\cite{asym}.  The slope of the wings is proportional to the lateral
field gradient $G_{zx}$. For $h \gg 2 R$, it is maximum at $x \simeq
0.39\, h$, where $h$ is the height between the sample surface and the
sphere center. At this location, the gradient is about $G_{zx} \approx
2.7 \, m_\text{sph} /h^4$. Since it is important to keep $h$ as large
as possible for stability purpose, the optimal $h$ is reached when
$\gamma G_{zx} R > \Omega$.
For our settings, this occurs at $h=1.8$~$\mu$m,
leading to slope of about $0.3$~GHz/$\mu$m. At this distance, the
maximum stray field of the sphere is about 140~G, a small variation
compared to the static perpendicular field of 1.72~T, ensuring that no
significant deformation of the SW modes profile is induced
\cite{klein08}.

We then turn to the variation of the amplitude of the f-MRFM signal as
a function of the position of the sphere in FIG.1b. The force acting
on the cantilever can be calculated as the vertical force exerted by
the tip on the sample, $F_z = G_{zz} \Delta M_z$, where $\Delta M_z$
is the variation of the sample magnetization induced by the FMR
resonance \cite{klein08}. The gradient $G_{zz}$ decays as the power
$1/x^5$, for large lateral displacement $x$. This decay ensures a
local detection.  Experimentally, the signal decreases by one order of
magnitude when the probe is displaced by 1.2~$\mu$m laterally.


\begin{figure}
\includegraphics[width=8.5cm]{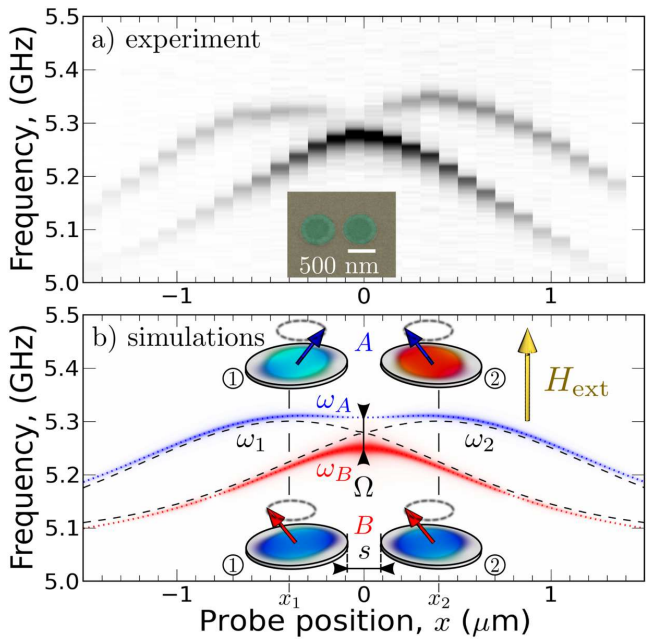}
\caption{a) Density plot of the experimental f-MRFM spectra as a
  function of the displacement $x$ of the sphere above a pair of 600
  nm Fe-V disks separated by $s=200$~nm. b) Predicted behavior by
  micromagnetic simulations. The upper mode (blue) corresponds to the
  anti-binding mode (A), while the lower (red) shows the binding mode
  (B). Insets: simulated precession profiles in each disk for modes
  (A) and (B) at the anti-crossing.  The dashed lines would be the
  individual modes of each disks in the absence of dynamical
  coupling.}
\label{600_200}
\end{figure}

We now discuss the same experiment above the pair of two 600 nm disks
separated by $s=200$~nm. The result is displayed on FIG.2a. Here $x=0$
corresponds to the middle of the pair. At each position $x$ of the
f-MRFM probe, we can see two modes. The upper branch has two frequency
maxima at $x_{1,2}=\mp 400$~nm, whose separation corresponds to the
center to center distance between disk $1$ and disk $2$. The two
maxima occur at slightly different frequencies, presumably due to a
small difference in diameter between the two disks. When the probe is
placed in between, $x_1 < x < x_2$, the two levels anti-cross, which
is a characteristic behavior of a coupled dynamics. Defining
$\omega_{1,2}$ as the frequencies of the two \textit{uncoupled} disks, the
collective frequencies follow:
\begin{equation}\label{eq:freq}
  \omega_{A,B} = \frac{\omega_1 + \omega_2}{2} \pm \sqrt{\left
      (\frac{\omega_1- \omega_2}{2} \right )^2 + \left
      (\frac{\Omega}{2} \right )^2 }  
\end{equation}
with ${\Omega}$ being the dynamical coupling strength. The two coupled
eigen-frequencies $\omega_{A,B}$ correspond respectively to the
anti-binding mode (A), where spins are precessing out-of-phase between
the two disks, and to the binding mode (B), where spins are precessing
in-phase \cite{naletov11}. In our f-MRFM experiment, $\omega_{1,2}$
both depend on $x$, see Eq.\ref{eq:omega}:
$\omega_{1,2}(x)=\omega_\text{FMR}+\gamma\{{H}_\text{sph,z}(x-x_{1,2})\}$. Using
these dependencies in Eq.\ref{eq:freq}, one can obtain an analytical
expression for the frequency difference $\omega_{A}(x)-\omega_{B}(x)$
observed in FIG.2. At $x=0$, when $\omega_1=\omega_2$, the splitting
$\omega_A-\omega_B$ exactly measures $\Omega$.
Using this analytical expression for the spatial dependence of the
splitting, we have fitted $\Omega/2 \pi = 50 \pm 5$~MHz. We emphasize
that this splitting is 2.5 times larger than the linewidth, found to be
$20$~MHz. Theoretically, the coupling $\Omega$ for the magneto-dipolar
interaction is defined by \cite{naletov11}
\begin{equation} \label{eq:Omega}
  \Omega^2 = 4 \gamma^2 h_{1,2}h_{2,1} \ .
\end{equation}
$h_{i,j}$ represents the cross depolarization field produced by the SW
in the $j$-th disk on the $i$-th disk ($i,j = 1,2$)
\cite{kostylev04,verba12}. It can be expressed as a function of the
cross depolarization tensor elements, which have an analytical
expression in the approximation of a uniform precession
\cite{tandon04}: $h_{i,j} = 2 \pi M_s ( \{N_{xx}^{i,j}\} +
\{N_{yy}^{i,j}\} )$. This formula reflects that the magneto-dipolar
interaction is anisotropic and thus, it induces an elliptical
precession in the two disks.  For the separation $s=200$~nm, a
numerical application yields $\{ N_{xx}^{1,2}\} \approx -2
\{N_{yy}^{1,2}\} \approx 0.0012$, which corresponds to a coupling
field of about 10~G between the two disks, or a coupling frequency
$\Omega/2 \pi = 56$~MHz, in very good agreement with the measured
value.

Another striking feature in FIG.2a is the strong variation of the
signal amplitude near the optimum coupling. We have explicitly plotted
in FIG.3a the amplitude of the f-MRFM signal as a function of the
lateral displacement $x$ of the probe, showing both the near
extinction of the anti-binding mode (A) and the strong enhancement of
the binding mode (B) near $x=0$. The ratio of hybridization in the two
coupled disks follows the expression:
\begin{equation} \label{eq:amp} \left . \frac{c_1}{c_2}
  \right|_{A,B}= \left ( \frac{(\omega_1 - \omega_2) \mp
    \sqrt{(\omega_1 - \omega_2)^2 + \Omega^2 }}{\Omega} \right )^{\mp 1}
\end{equation}
Introducing the spatial dependence of $\omega_{1,2}$ described by
Eq.\ref{eq:omega} in Eq.\ref{eq:amp} we can calculate the total force
$ F_z \propto P \left [ c_1^2 \, G_{zz}(x-x_1) + c_2^2\, G_{zz}(x-x_2)
\right ]$ acting on the cantilever. The power efficiency $P = | c_1 +
c_2 |^2 h_\text{rf}^2$ is proportional to the overlap integral between
the uniform rf field and the collective SW mode (the vector sum of the
transverse magnetization in the two disks) \cite{naletov11}.
Using Eq.\ref{eq:amp}, the dependence on $x$ of the force produced by
the binding and anti-binding mode gives the continuous lines shown in
FIG.3a. The difference between the two curves comes mainly from the
selection rules defined in $P$. We find that at the optimum coupling
(when $\omega_1$ and $\omega_2$ cross), the anti-binding mode (A) in
Eq.\ref{eq:amp} has $c_1 = - c_2$ , \textit{i.e.}, a precession with
equal hybridization weight between the two disks and out-of-phase. The
overlap with the uniform rf field excitation is thus zero at $x=0$,
leading to a vanishing amplitude. In contrast, the binding mode (B)
has $c_1 = + c_2$ at the anti-crossing, \textit{i.e.}, a precession
with equal hybridization weight too, but now in-phase between the two
disks. It represents and enhancement of the absorbed power by a factor
of $2^2$ compared to the amplitude in one disk. 

We then study the effect the magneto-dipolar coupling on the linewidth
of the collective mode. We observe that the linewidth does not change
much with tuning and the observed variation with $x$ is below the 5\%
range. At the optimal tuning $x=0$, the linewidth measured is $\Delta
f=22.3\pm0.5$~MHz (see FIG.3b) and it becomes slightly larger $\Delta
f=23.1\pm0.5$~MHz at the maximum detuning $x=x_{1,2}$. For comparison
we have displayed in FIG.3c the linewidth observed above the single
disk, whose value $\Delta f=21.4\pm0.5$~MHz.  A small increase of the
ratio $\Delta f / f$ is indeed expected for the dynamically coupled
modes. This comes from the fact that this ratio is equal to $\Delta f
/ f= \alpha (H_x + H_y)/\sqrt{H_x H_y} $, where $\alpha$ is the
Gilbert damping, and $H_x$ and $H_y$ represent the two stiffness
fields which characterize the torque exerted on the magnetization when
it is tipped along the $x$- or $y$-axis \cite{gurevich96}.  The degree
of hybridization as well as the nature of the mode (A or B) change the
values and signs of $H_x$ and $H_y$. For the binding mode, the
magneto-dipolar coupling generates an elliptical precession whose long
axis is along the two disks axis. The induced ellipticity
$\mathcal{E}$ is maximum at the anti-crossing ($x=0$), with an
amplitude $\mathcal{E} =
\frac{\beta-1}{\beta+1}\frac{\Omega}{\omega_B} \approx 3\%$, with
$\beta = \{N_{xx}^{1,2}\}/\{N_{yy}^{1,2}\}\approx -2$. An increase of
ellipticity induces an increase of the linewdith, a behavior which is
consistent with the small additional broadening measured in our
experiment.

The analytical model used above to analyze the data assumes a uniform
magnetization throughout the magnetic body. To take more precisely
into account the 3D texture of the magnetization and the static
deformation induced by the probe, we have also calculated the
eigen-frequencies of the two lowest energy modes as a function of $x$
using SpinFlow3D, a finite element solver developed by In Silicio
\cite{SpinFlow3D}. The disks are discretized with a mesh size of 10~nm
using a Delaunay mesh construction. At each position of the probe, we
first calculate the equilibrium configuration in the disks. The
Arnoldi algorithm is then used to compute the lowest eigen-values of
the problem as well as the associated eigen-vectors. The result is
represented in red and blue in FIG.2b for the two lowest energy
modes. The precession patterns associated to each mode at the
anti-crossing are shown in inset. In this color representation, the
hue indicates the phase (or direction) of the oscillating
magnetization, while the brightness indicates its amplitude. The
simulation results confirm very nicely the interpretation made above
in terms of amplitude and peak position.

\begin{figure}
\includegraphics[width=8.5cm]{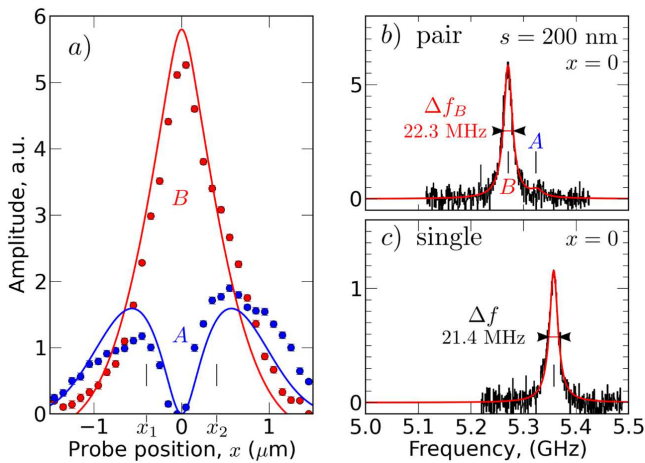}
\caption{a) Variation of the amplitude of the binding (red) and
  anti-binding (blue) resonances as a function of the lateral position
  of the probe for the two disks separated by 200~nm. The solid lines
  correspond to the behavior following from Eq.\ref{eq:amp} (see
  text). b) Linewidth of the binding mode for the same pair at the
  tuning position ($x=0$) c) Comparison with the measurement of the
  linewidth above a single disk. }
\label{amp_width}
\end{figure}

We have then repeated the same procedure on the two other pairs of
disks, with larger edge to edge separation $s$. The strength of the
dynamical coupling measured by f-MRFM is plotted as a function of $s$
in FIG.4. The main results is that, with our experimental parameters,
$s$ needs to be less than the diameter of the disks in order to have
$\Omega$ larger than the linewidth $\Delta f$. The data are plotted
along with the analytical prediction (continuous line) and the
simulations (dashed line with small dots). We observe an excellent
overall agreement between the three sets of results, which all exhibit
a similar decay with $s$ (not a simple polynomial law
\cite{sukhostavets11}). Still, the experimental points are
systematically slightly below the theoretical expectation. This could
be explained by the fact that the disks are slightly smaller than
their nominal value (\textit{e.g.}, due to some oxidation at their
periphery), or that the true separation between the disks is slightly
larger than expected, which we have represented on the graph by the
horizontal error bars. The agreement between the analytical model and
the simulation is very good until $s=0.1$~$\mu$m. The discrepancy for
very small $s$ is due to a significative change in the static magnetic
texture. These changes are not taken into account by the analytical
model. The effect of the static coupling is to produce a static
magnetization along the $x$-direction. As shown by the simulations,
this deformation enhances the strength of the dynamical coupling.


\begin{figure}
\includegraphics[width=8.5cm]{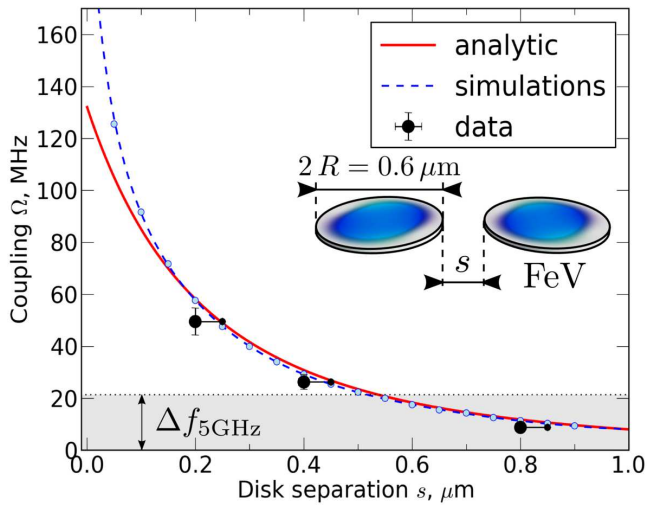}
\caption{Coupling strength as a function of the separation $s$ between
  two disks. The plot compares the experimental findings to the
  predicted amplitude of the magneto-dipolar interaction either
  analytically (continuous line) or by micromagnetic simulations
  (small dots, dashed line is a guide to the eye).}
\label{coupling}
\end{figure}

In conclusion, we have shown that f-MRFM enables a detailed
investigation of the dynamical dipolar coupling between two nearby
magnetic objects, owing to the possibility of the technique to study
both the tuned and detuned regime on the same object. It has been
applied to study the collective SW dynamics in pairs nano-disks of
Fe-V, an ultra-low damping material. Several signatures of the
collective behavior have been experimentally evidenced and
quantitatively explained: the anti-crossing, the hybridization of the
modes and the effects on the linewidth. Moreover, we have found that
in order to have a frequency splitting larger than the linewidth of
the modes, the edge to edge separation between our disks has to be
smaller than their diameter, due to the fast decay of the
magneto-dipolar interaction. We believe that our method of local
characterization of the dipolar coupling will be very useful to the
field of magnonics.

This research work was partially supported by the French Grants VOICE
ANR-09-NANO-006-01 and MARVEL ANR-2010-JCJC-0410-01.


\begin{thebibliography}{0}%
\makeatletter
\providecommand \@ifxundefined [1]{%
 \@ifx{#1\undefined}
}%
\providecommand \@ifnum [1]{%
 \ifnum #1\expandafter \@firstoftwo
 \else \expandafter \@secondoftwo
 \fi
}%
\providecommand \@ifx [1]{%
 \ifx #1\expandafter \@firstoftwo
 \else \expandafter \@secondoftwo
 \fi
}%
\providecommand \natexlab [1]{#1}%
\providecommand \enquote  [1]{``#1''}%
\providecommand \bibnamefont  [1]{#1}%
\providecommand \bibfnamefont [1]{#1}%
\providecommand \citenamefont [1]{#1}%
\providecommand \href@noop [0]{\@secondoftwo}%
\providecommand \href [0]{\begingroup \@sanitize@url \@href}%
\providecommand \@href[1]{\@@startlink{#1}\@@href}%
\providecommand \@@href[1]{\endgroup#1\@@endlink}%
\providecommand \@sanitize@url [0]{\catcode `\\12\catcode `\$12\catcode
  `\&12\catcode `\#12\catcode `\^12\catcode `\_12\catcode `\%12\relax}%
\providecommand \@@startlink[1]{}%
\providecommand \@@endlink[0]{}%
\providecommand \url  [0]{\begingroup\@sanitize@url \@url }%
\providecommand \@url [1]{\endgroup\@href {#1}{\urlprefix }}%
\providecommand \urlprefix  [0]{URL }%
\providecommand \Eprint [0]{\href }%
\providecommand \doibase [0]{http://dx.doi.org/}%
\providecommand \selectlanguage [0]{\@gobble}%
\providecommand \bibinfo  [0]{\@secondoftwo}%
\providecommand \bibfield  [0]{\@secondoftwo}%
\providecommand \translation [1]{[#1]}%
\providecommand \BibitemOpen [0]{}%
\providecommand \bibitemStop [0]{}%
\providecommand \bibitemNoStop [0]{.\EOS\space}%
\providecommand \EOS [0]{\spacefactor3000\relax}%
\providecommand \BibitemShut  [1]{\csname bibitem#1\endcsname}%
\let\auto@bib@innerbib\@empty
\end{thebibliography}%


\begin{thebibliography}{10}%
\makeatletter
\providecommand \@ifxundefined [1]{%
 \ifx #1\undefined \expandafter \@firstoftwo
 \else \expandafter \@secondoftwo
\fi
}%
\providecommand \@ifnum [1]{%
 \ifnum #1\expandafter \@firstoftwo
 \else \expandafter \@secondoftwo
\fi
}%
\providecommand \enquote [1]{``#1''}%
\providecommand \bibnamefont  [1]{#1}%
\providecommand \bibfnamefont [1]{#1}%
\providecommand \citenamefont [1]{#1}%
\providecommand\href[0]{\@sanitize\@href}%
\providecommand\@href[1]{\endgroup\@@startlink{#1}\endgroup\@@href}%
\providecommand\@@href[1]{#1\@@endlink}%
\providecommand \@sanitize [0]{\begingroup\catcode`\&12\catcode`\#12\relax}%
\@ifxundefined \pdfoutput {\@firstoftwo}{%
 \@ifnum{\z@=\pdfoutput}{\@firstoftwo}{\@secondoftwo}%
}{%
 \providecommand\@@startlink[1]{\leavevmode}%
 \providecommand\@@endlink[0]{}%
}{%
 \providecommand\@@startlink[1]{%
  \leavevmode
  \pdfstartlink
   attr{/Border[0 0 1 ]/H/I/C[0 1 1]}%
   user{/Subtype/Link/A<</Type/Action/S/URI/URI(#1)>>}%
  \relax
 }%
 \providecommand\@@endlink[0]{\pdfendlink}%
}%
\providecommand \url  [0]{\begingroup\@sanitize \@url }%
\providecommand \@url [1]{\endgroup\@href {#1}{\urlprefix}}%
\providecommand \urlprefix [0]{URL }%
\providecommand \Eprint[0]{\href }%
\@ifxundefined \urlstyle {%
  \providecommand \doi [1]{doi:\discretionary{}{}{}#1}%
}{%
  \providecommand \doi [0]{doi:\discretionary{}{}{}\begingroup
  \urlstyle{rm}\Url }%
}%
\providecommand \doibase [0]{http://dx.doi.org/}%
\providecommand \Doi[1]{\href{\doibase#1}}%
\providecommand \bibAnnote [3]{%
  \BibitemShut{#1}%
  \begin{quotation}\noindent
    \textsc{Key:}\ #2\\\textsc{Annotation:}\ #3%
  \end{quotation}%
}%
\providecommand \bibAnnoteFile [2]{%
  \IfFileExists{#2}{\bibAnnote {#1} {#2} {\input{#2}}}{}%
}%
\providecommand \typeout [0]{\immediate \write \m@ne }%
\providecommand \selectlanguage [0]{\@gobble}%
\providecommand \bibinfo [0]{\@secondoftwo}%
\providecommand \bibfield [0]{\@secondoftwo}%
\providecommand \translation [1]{[#1]}%
\providecommand \BibitemOpen[0]{}%
\providecommand \bibitemStop [0]{}%
\providecommand \bibitemNoStop [0]{.\EOS\space}%
\providecommand \EOS [0]{\spacefactor3000\relax}%
\providecommand \BibitemShut [1]{\csname bibitem#1\endcsname}%
\bibitem{chou06}%
  \BibitemOpen
  \bibfield{author}{%
  \bibinfo {author} {\bibfnamefont{K.~W.}\ \bibnamefont{Chou}}, \bibinfo
  {author} {\bibfnamefont{A.}~\bibnamefont{Puzic}}, \bibinfo {author}
  {\bibfnamefont{H.}~\bibnamefont{Stoll}}, \bibinfo {author}
  {\bibfnamefont{G.}~\bibnamefont{Sch\"{u}tz}}, \bibinfo {author}
  {\bibfnamefont{B.~V.}\ \bibnamefont{Waeyenberge}}, \bibinfo {author}
  {\bibfnamefont{T.}~\bibnamefont{Tyliszczak}}, \bibinfo {author}
  {\bibfnamefont{K.}~\bibnamefont{Rott}}, \bibinfo {author}
  {\bibfnamefont{G.}~\bibnamefont{Reiss}}, \bibinfo {author}
  {\bibfnamefont{H.}~\bibnamefont{Br\"{u}ckl}}, \bibinfo {author}
  {\bibfnamefont{I.}~\bibnamefont{Neudecker}}, \bibinfo {author}
  {\bibfnamefont{D.}~\bibnamefont{Weiss}},\ and\ \bibinfo {author}
  {\bibfnamefont{C.~H.}\ \bibnamefont{Back}},\ }%
  \bibfield{journal}{%
  \Doi{10.1063/1.2173630}{\bibinfo {journal} {J. Appl. Phys.}}\ }%
  \textbf{\bibinfo {volume} {99}},\ \bibinfo {eid} {08F305} (\bibinfo {year}
  {2006})%
  \bibAnnoteFile{NoStop}{chou06}%
\bibitem{loubens07a}%
  \BibitemOpen
  \bibfield{author}{%
  \bibinfo {author} {\bibfnamefont{G.}~\bibnamefont{de~Loubens}}, \bibinfo
  {author} {\bibfnamefont{V.~V.}\ \bibnamefont{Naletov}}, \bibinfo {author}
  {\bibfnamefont{M.}~\bibnamefont{Viret}}, \bibinfo {author}
  {\bibfnamefont{O.}~\bibnamefont{Klein}}, \bibinfo {author}
  {\bibfnamefont{H.}~\bibnamefont{Hurdequint}}, \bibinfo {author}
  {\bibfnamefont{J.}~\bibnamefont{{Ben Youssef}}}, \bibinfo {author}
  {\bibfnamefont{F.}~\bibnamefont{Boust}},\ and\ \bibinfo {author}
  {\bibfnamefont{N.}~\bibnamefont{Vukadinovic}},\ }%
  \bibfield{journal}{%
  \bibinfo {journal} {J. Appl. Phys.}\ }%
  \textbf{\bibinfo {volume} {101}},\ \bibinfo {pages} {09F514} (\bibinfo {year}
  {2007})%
  \bibAnnoteFile{NoStop}{loubens07a}%
\bibitem{gubbiotti08}%
  \BibitemOpen
  \bibfield{author}{%
  \bibinfo {author} {\bibfnamefont{G.}~\bibnamefont{Gubbiotti}}, \bibinfo
  {author} {\bibfnamefont{M.}~\bibnamefont{Madami}}, \bibinfo {author}
  {\bibfnamefont{S.}~\bibnamefont{Tacchi}}, \bibinfo {author}
  {\bibfnamefont{G.}~\bibnamefont{Carlotti}}, \bibinfo {author}
  {\bibfnamefont{H.}~\bibnamefont{Tanigawa}},\ and\ \bibinfo {author}
  {\bibfnamefont{T.}~\bibnamefont{Ono}},\ }%
  \bibfield{journal}{%
  \Doi{10.1088/0022-3727/41/13/134023}{\bibinfo {journal} {J. Phys. D: Appl.
  Phys.}}\ }%
  \textbf{\bibinfo {volume} {41}},\ \bibinfo {pages} {134023} (\bibinfo {year}
  {2008})%
  \bibAnnoteFile{NoStop}{gubbiotti08}%
\bibitem{awad10a}%
  \BibitemOpen
  \bibfield{author}{%
  \bibinfo {author} {\bibfnamefont{A.~A.}\ \bibnamefont{Awad}}, \bibinfo
  {author} {\bibfnamefont{G.~R.}\ \bibnamefont{Aranda}}, \bibinfo {author}
  {\bibfnamefont{D.}~\bibnamefont{Dieleman}}, \bibinfo {author}
  {\bibfnamefont{K.~Y.}\ \bibnamefont{Guslienko}}, \bibinfo {author}
  {\bibfnamefont{G.~N.}\ \bibnamefont{Kakazei}}, \bibinfo {author}
  {\bibfnamefont{B.~A.}\ \bibnamefont{Ivanov}},\ and\ \bibinfo {author}
  {\bibfnamefont{F.~G.}\ \bibnamefont{Aliev}},\ }%
  \bibfield{journal}{%
  \Doi{10.1063/1.3495774}{\bibinfo {journal} {Appl. Phys. Lett.}}\ }%
  \textbf{\bibinfo {volume} {97}},\ \bibinfo {eid} {132501} (\bibinfo {year}
  {2010})%
  \bibAnnoteFile{NoStop}{awad10a}%
\bibitem{jung10}%
  \BibitemOpen
  \bibfield{author}{%
  \bibinfo {author} {\bibfnamefont{H.}~\bibnamefont{Jung}}, \bibinfo {author}
  {\bibfnamefont{Y.-S.}\ \bibnamefont{Yu}}, \bibinfo {author}
  {\bibfnamefont{K.-S.}\ \bibnamefont{Lee}}, \bibinfo {author}
  {\bibfnamefont{M.-Y.}\ \bibnamefont{Im}}, \bibinfo {author}
  {\bibfnamefont{P.}~\bibnamefont{Fischer}}, \bibinfo {author}
  {\bibfnamefont{L.}~\bibnamefont{Bocklage}}, \bibinfo {author}
  {\bibfnamefont{A.}~\bibnamefont{Vogel}}, \bibinfo {author}
  {\bibfnamefont{M.}~\bibnamefont{Bolte}}, \bibinfo {author}
  {\bibfnamefont{G.}~\bibnamefont{Meier}},\ and\ \bibinfo {author}
  {\bibfnamefont{S.-K.}\ \bibnamefont{Kim}},\ }%
  \bibfield{journal}{%
  \Doi{10.1063/1.3517496}{\bibinfo {journal} {Appl. Phys. Lett.}}\ }%
  \textbf{\bibinfo {volume} {97}},\ \bibinfo {eid} {222502} (\bibinfo {year}
  {2010})%
  \bibAnnoteFile{NoStop}{jung10}%
\bibitem{vogel10}%
  \BibitemOpen
  \bibfield{author}{%
  \bibinfo {author} {\bibfnamefont{A.}~\bibnamefont{Vogel}}, \bibinfo {author}
  {\bibfnamefont{A.}~\bibnamefont{Drews}}, \bibinfo {author}
  {\bibfnamefont{T.}~\bibnamefont{Kamionka}}, \bibinfo {author}
  {\bibfnamefont{M.}~\bibnamefont{Bolte}},\ and\ \bibinfo {author}
  {\bibfnamefont{G.}~\bibnamefont{Meier}},\ }%
  \bibfield{journal}{%
  \Doi{10.1103/PhysRevLett.105.037201}{\bibinfo {journal} {Phys. Rev. Lett.}}\
  }%
  \textbf{\bibinfo {volume} {105}},\ \bibinfo {pages} {037201} (\bibinfo {year}
  {2010})%
  \bibAnnoteFile{NoStop}{vogel10}%
\bibitem{sugimoto11}%
  \BibitemOpen
  \bibfield{author}{%
  \bibinfo {author} {\bibfnamefont{S.}~\bibnamefont{Sugimoto}}, \bibinfo
  {author} {\bibfnamefont{Y.}~\bibnamefont{Fukuma}}, \bibinfo {author}
  {\bibfnamefont{S.}~\bibnamefont{Kasai}}, \bibinfo {author}
  {\bibfnamefont{T.}~\bibnamefont{Kimura}}, \bibinfo {author}
  {\bibfnamefont{A.}~\bibnamefont{Barman}},\ and\ \bibinfo {author}
  {\bibfnamefont{Y.}~\bibnamefont{Otani}},\ }%
  \bibfield{journal}{%
  \Doi{10.1103/PhysRevLett.106.197203}{\bibinfo {journal} {Phys. Rev. Lett.}}\
  }%
  \textbf{\bibinfo {volume} {106}},\ \bibinfo {pages} {197203} (\bibinfo {year}
  {2011})%
  \bibAnnoteFile{NoStop}{sugimoto11}%
\bibitem{ulrichs11}%
  \BibitemOpen
  \bibfield{author}{%
  \bibinfo {author} {\bibfnamefont{H.}~\bibnamefont{Ulrichs}}, \bibinfo
  {author} {\bibfnamefont{V.~E.}\ \bibnamefont{Demidov}}, \bibinfo {author}
  {\bibfnamefont{S.~O.}\ \bibnamefont{Demokritov}}, \bibinfo {author}
  {\bibfnamefont{A.~V.}\ \bibnamefont{Ognev}}, \bibinfo {author}
  {\bibfnamefont{M.~E.}\ \bibnamefont{Stebliy}}, \bibinfo {author}
  {\bibfnamefont{L.~A.}\ \bibnamefont{Chebotkevich}},\ and\ \bibinfo {author}
  {\bibfnamefont{A.~S.}\ \bibnamefont{Samardak}},\ }%
  \bibfield{journal}{%
  \Doi{10.1103/PhysRevB.83.184403}{\bibinfo {journal} {Phys. Rev. B}}\ }%
  \textbf{\bibinfo {volume} {83}},\ \bibinfo {pages} {184403} (\bibinfo {year}
  {2011})%
  \bibAnnoteFile{NoStop}{ulrichs11}%
\bibitem{kruglyak10}%
  \BibitemOpen
  \bibfield{author}{%
  \bibinfo {author} {\bibfnamefont{V.~V.}\ \bibnamefont{Kruglyak}}, \bibinfo
  {author} {\bibfnamefont{P.~S.}\ \bibnamefont{Keatley}}, \bibinfo {author}
  {\bibfnamefont{A.}~\bibnamefont{Neudert}}, \bibinfo {author}
  {\bibfnamefont{R.~J.}\ \bibnamefont{Hicken}}, \bibinfo {author}
  {\bibfnamefont{J.~R.}\ \bibnamefont{Childress}},\ and\ \bibinfo {author}
  {\bibfnamefont{J.~A.}\ \bibnamefont{Katine}},\ }%
  \bibfield{journal}{%
  \Doi{10.1103/PhysRevLett.104.027201}{\bibinfo {journal} {Phys. Rev. Lett.}}\
  }%
  \textbf{\bibinfo {volume} {104}},\ \bibinfo {pages} {027201} (\bibinfo {year}
  {2010})%
  \bibAnnoteFile{NoStop}{kruglyak10}%
\bibitem{belanovsky12}%
  \BibitemOpen
  \bibfield{author}{%
  \bibinfo {author} {\bibfnamefont{A.~D.}\ \bibnamefont{Belanovsky}}, \bibinfo
  {author} {\bibfnamefont{N.}~\bibnamefont{Locatelli}}, \bibinfo {author}
  {\bibfnamefont{P.~N.}\ \bibnamefont{Skirdkov}}, \bibinfo {author}
  {\bibfnamefont{F.~A.}\ \bibnamefont{Araujo}}, \bibinfo {author}
  {\bibfnamefont{J.}~\bibnamefont{Grollier}}, \bibinfo {author}
  {\bibfnamefont{K.~A.}\ \bibnamefont{Zvezdin}}, \bibinfo {author}
  {\bibfnamefont{V.}~\bibnamefont{Cros}},\ and\ \bibinfo {author}
  {\bibfnamefont{A.~K.}\ \bibnamefont{Zvezdin}},\ }%
  \bibfield{journal}{%
  \Doi{10.1103/PhysRevB.85.100409}{\bibinfo {journal} {Phys. Rev. B}}\ }%
  \textbf{\bibinfo {volume} {85}},\ \bibinfo {pages} {100409} (\bibinfo {year}
  {2012})%
  \bibAnnoteFile{NoStop}{belanovsky12}%
\bibitem{kruglyak10a}%
  \BibitemOpen
  \bibfield{author}{%
  \bibinfo {author} {\bibfnamefont{V.~V.}\ \bibnamefont{Kruglyak}}, \bibinfo
  {author} {\bibfnamefont{S.~O.}\ \bibnamefont{Demokritov}},\ and\ \bibinfo
  {author} {\bibfnamefont{D.}~\bibnamefont{Grundler}},\ }%
  \bibfield{journal}{%
  \bibinfo {journal} {Journal of Physics D: Applied Physics}\ }%
  \textbf{\bibinfo {volume} {43}},\ \bibinfo {pages} {264001} (\bibinfo {year}
  {2010})%
  \bibAnnoteFile{NoStop}{kruglyak10a}%
\bibitem{karenowska12}%
  \BibitemOpen
  \bibfield{author}{%
  \bibinfo {author} {\bibfnamefont{A.~D.}\ \bibnamefont{Karenowska}}, \bibinfo
  {author} {\bibfnamefont{J.~F.}\ \bibnamefont{Gregg}}, \bibinfo {author}
  {\bibfnamefont{V.~S.}\ \bibnamefont{Tiberkevich}}, \bibinfo {author}
  {\bibfnamefont{A.~N.}\ \bibnamefont{Slavin}}, \bibinfo {author}
  {\bibfnamefont{A.~V.}\ \bibnamefont{Chumak}}, \bibinfo {author}
  {\bibfnamefont{A.~A.}\ \bibnamefont{Serga}},\ and\ \bibinfo {author}
  {\bibfnamefont{B.}~\bibnamefont{Hillebrands}},\ }%
  \bibfield{journal}{%
  \Doi{10.1103/PhysRevLett.108.015505}{\bibinfo {journal} {Phys. Rev. Lett.}}\
  }%
  \textbf{\bibinfo {volume} {108}},\ \bibinfo {pages} {015505} (\bibinfo {year}
  {2012})%
  \bibAnnoteFile{NoStop}{karenowska12}%
\bibitem{kubo11}%
  \BibitemOpen
  \bibfield{author}{%
  \bibinfo {author} {\bibfnamefont{Y.}~\bibnamefont{Kubo}}, \bibinfo {author}
  {\bibfnamefont{C.}~\bibnamefont{Grezes}}, \bibinfo {author}
  {\bibfnamefont{A.}~\bibnamefont{Dewes}}, \bibinfo {author}
  {\bibfnamefont{T.}~\bibnamefont{Umeda}}, \bibinfo {author}
  {\bibfnamefont{J.}~\bibnamefont{Isoya}}, \bibinfo {author}
  {\bibfnamefont{H.}~\bibnamefont{Sumiya}}, \bibinfo {author}
  {\bibfnamefont{N.}~\bibnamefont{Morishita}}, \bibinfo {author}
  {\bibfnamefont{H.}~\bibnamefont{Abe}}, \bibinfo {author}
  {\bibfnamefont{S.}~\bibnamefont{Onoda}}, \bibinfo {author}
  {\bibfnamefont{T.}~\bibnamefont{Ohshima}}, \bibinfo {author}
  {\bibfnamefont{V.}~\bibnamefont{Jacques}}, \bibinfo {author}
  {\bibfnamefont{A.}~\bibnamefont{Dr\'eau}}, \bibinfo {author}
  {\bibfnamefont{J.-F.}\ \bibnamefont{Roch}}, \bibinfo {author}
  {\bibfnamefont{I.}~\bibnamefont{Diniz}}, \bibinfo {author}
  {\bibfnamefont{A.}~\bibnamefont{Auffeves}}, \bibinfo {author}
  {\bibfnamefont{D.}~\bibnamefont{Vion}}, \bibinfo {author}
  {\bibfnamefont{D.}~\bibnamefont{Esteve}},\ and\ \bibinfo {author}
  {\bibfnamefont{P.}~\bibnamefont{Bertet}},\ }%
  \bibfield{journal}{%
  \Doi{10.1103/PhysRevLett.107.220501}{\bibinfo {journal} {Phys. Rev. Lett.}}\
  }%
  \textbf{\bibinfo {volume} {107}},\ \bibinfo {pages} {220501} (\bibinfo {year}
  {2011})%
  \bibAnnoteFile{NoStop}{kubo11}%
\bibitem{bonell09}%
  \BibitemOpen
  \bibfield{author}{%
  \bibinfo {author} {\bibfnamefont{F.}~\bibnamefont{Bonell}}, \bibinfo {author}
  {\bibfnamefont{S.}~\bibnamefont{Andrieu}}, \bibinfo {author}
  {\bibfnamefont{F.}~\bibnamefont{Bertran}}, \bibinfo {author}
  {\bibfnamefont{P.}~\bibnamefont{Lefevre}}, \bibinfo {author}
  {\bibfnamefont{A.}~\bibnamefont{Ibrahimi}}, \bibinfo {author}
  {\bibfnamefont{E.}~\bibnamefont{Snoeck}}, \bibinfo {author}
  {\bibfnamefont{C.-V.}\ \bibnamefont{Tiusan}},\ and\ \bibinfo {author}
  {\bibfnamefont{F.}~\bibnamefont{Montaigne}},\ }%
  \bibfield{journal}{%
  \Doi{10.1109/TMAG.2009.2022644}{\bibinfo {journal} {IEEE Trans. Magn.}}\ }%
  \textbf{\bibinfo {volume} {45}},\ \bibinfo {pages} {3467 } (\bibinfo {year}
  {2009})%
  \bibAnnoteFile{NoStop}{bonell09}%
\bibitem{mitsuzuka12}%
  \BibitemOpen
  \bibfield{author}{%
  \bibinfo {author} {\bibfnamefont{K.}~\bibnamefont{Mitsuzuka}}, \bibinfo
  {author} {\bibfnamefont{D.}~\bibnamefont{Lacour}}, \bibinfo {author}
  {\bibfnamefont{M.}~\bibnamefont{Hehn}}, \bibinfo {author}
  {\bibfnamefont{S.}~\bibnamefont{Andrieu}},\ and\ \bibinfo {author}
  {\bibfnamefont{F.}~\bibnamefont{Montaigne}},\ }%
  \bibfield{journal}{%
  \Doi{10.1063/1.4711219}{\bibinfo {journal} {Appl. Phys. Lett.}}\ }%
  \textbf{\bibinfo {volume} {100}},\ \bibinfo {eid} {192406} (\bibinfo {year}
  {2012})%
  \bibAnnoteFile{NoStop}{mitsuzuka12}%
\bibitem{angle}%
  \BibitemOpen
  \bibinfo {note} {In fact, the external field is tilted by $\theta_H \simeq
  2^\circ$ in the $x$-direction. Our setup does not allow in-situ adjustment of
  this small misalignment. Although it can be easily integrated in a complete
  analysis \cite{klein08}, this point will be neglected for the sake of
  simplicity as it brings minor correction.}%
  \bibAnnoteFile{Stop}{angle}%
\bibitem{klein08}%
  \BibitemOpen
  \bibfield{author}{%
  \bibinfo {author} {\bibfnamefont{O.}~\bibnamefont{Klein}}, \bibinfo {author}
  {\bibfnamefont{G.}~\bibnamefont{de~Loubens}}, \bibinfo {author}
  {\bibfnamefont{V.~V.}\ \bibnamefont{Naletov}}, \bibinfo {author}
  {\bibfnamefont{F.}~\bibnamefont{Boust}}, \bibinfo {author}
  {\bibfnamefont{T.}~\bibnamefont{Guillet}}, \bibinfo {author}
  {\bibfnamefont{H.}~\bibnamefont{Hurdequint}}, \bibinfo {author}
  {\bibfnamefont{A.}~\bibnamefont{Leksikov}}, \bibinfo {author}
  {\bibfnamefont{A.~N.}\ \bibnamefont{Slavin}}, \bibinfo {author}
  {\bibfnamefont{V.~S.}\ \bibnamefont{Tiberkevich}},\ and\ \bibinfo {author}
  {\bibfnamefont{N.}~\bibnamefont{Vukadinovic}},\ }%
  \bibfield{journal}{%
  \Doi{10.1103/PhysRevB.78.144410}{\bibinfo {journal} {Phys. Rev. B}}\ }%
  \textbf{\bibinfo {volume} {78}},\ \bibinfo {eid} {144410} (\bibinfo {year}
  {2008})%
  \bibAnnoteFile{NoStop}{klein08}%
\bibitem{lee10}%
  \BibitemOpen
  \bibfield{author}{%
  \bibinfo {author} {\bibfnamefont{I.}~\bibnamefont{Lee}}, \bibinfo {author}
  {\bibfnamefont{Y.}~\bibnamefont{Obukhov}}, \bibinfo {author}
  {\bibfnamefont{G.}~\bibnamefont{Xiang}}, \bibinfo {author}
  {\bibfnamefont{A.}~\bibnamefont{Hauser}}, \bibinfo {author}
  {\bibfnamefont{F.}~\bibnamefont{Yang}}, \bibinfo {author}
  {\bibfnamefont{P.}~\bibnamefont{Banerjee}}, \bibinfo {author}
  {\bibfnamefont{D.}~\bibnamefont{Pelekhov}},\ and\ \bibinfo {author}
  {\bibfnamefont{P.}~\bibnamefont{Hammel}},\ }%
  \bibfield{journal}{%
  \bibinfo {journal} {Nature (London)}\ }%
  \textbf{\bibinfo {volume} {466}},\ \bibinfo {pages} {845} (\bibinfo {year}
  {2010})%
  \bibAnnoteFile{NoStop}{lee10}%
\bibitem{Bessel}%
  \BibitemOpen
  \bibinfo {note} {$ \{ \bm{H}_\text{sph} \}= \int_{S} \frac{J_0(\kappa_{0,0}
  \sqrt{x^2 + y^2}/ R)^2 \bm{H}_\text{sph}(x,y)}{\pi J_1(\kappa_{0,0})^2 R^2}
  \, dxdy $ where $S$ is the area of the disk, $J_\ell$ is the $\ell$-th order
  Bessel function and $\kappa_{\ell, n}$ is its $n$-th order root
  \cite{klein08}.}%
  \bibAnnoteFile{Stop}{Bessel}%
\bibitem{asym}%
  \BibitemOpen
  \bibinfo {note} {The asymmetries of the bell-shape curve (maximum of
  frequency shifted with respect to $x=0$ and to the maximum of amplitude, and
  different slopes in the wings) are due to the misalignment $\theta_H$ noted
  in \cite{angle}.}%
  \bibAnnoteFile{Stop}{asym}%
\bibitem{naletov11}%
  \BibitemOpen
  \bibfield{author}{%
  \bibinfo {author} {\bibfnamefont{V.~V.}\ \bibnamefont{Naletov}}, \bibinfo
  {author} {\bibfnamefont{G.}~\bibnamefont{de~Loubens}}, \bibinfo {author}
  {\bibfnamefont{G.}~\bibnamefont{Albuquerque}}, \bibinfo {author}
  {\bibfnamefont{S.}~\bibnamefont{Borlenghi}}, \bibinfo {author}
  {\bibfnamefont{V.}~\bibnamefont{Cros}}, \bibinfo {author}
  {\bibfnamefont{G.}~\bibnamefont{Faini}}, \bibinfo {author}
  {\bibfnamefont{J.}~\bibnamefont{Grollier}}, \bibinfo {author}
  {\bibfnamefont{H.}~\bibnamefont{Hurdequint}}, \bibinfo {author}
  {\bibfnamefont{N.}~\bibnamefont{Locatelli}}, \bibinfo {author}
  {\bibfnamefont{B.}~\bibnamefont{Pigeau}}, \bibinfo {author}
  {\bibfnamefont{A.~N.}\ \bibnamefont{Slavin}}, \bibinfo {author}
  {\bibfnamefont{V.~S.}\ \bibnamefont{Tiberkevich}}, \bibinfo {author}
  {\bibfnamefont{C.}~\bibnamefont{Ulysse}}, \bibinfo {author}
  {\bibfnamefont{T.}~\bibnamefont{Valet}},\ and\ \bibinfo {author}
  {\bibfnamefont{O.}~\bibnamefont{Klein}},\ }%
  \bibfield{journal}{%
  \Doi{10.1103/PhysRevB.84.224423}{\bibinfo {journal} {Phys. Rev. B}}\ }%
  \textbf{\bibinfo {volume} {84}},\ \bibinfo {pages} {224423} (\bibinfo {year}
  {2011})%
  \bibAnnoteFile{NoStop}{naletov11}%
\bibitem{kostylev04}%
  \BibitemOpen
  \bibfield{author}{%
  \bibinfo {author} {\bibfnamefont{M.~P.}\ \bibnamefont{Kostylev}}, \bibinfo
  {author} {\bibfnamefont{A.~A.}\ \bibnamefont{Stashkevich}}, \bibinfo {author}
  {\bibfnamefont{N.~A.}\ \bibnamefont{Sergeeva}},\ and\ \bibinfo {author}
  {\bibfnamefont{Y.}~\bibnamefont{Roussign\'e}},\ }%
  \bibfield{journal}{%
  \Doi{10.1016/j.jmmm.2003.11.400}{\bibinfo {journal} {J. Magn. Magn. Mater.}}\
  }%
  \textbf{\bibinfo {volume} {278}},\ \bibinfo {pages} {397} (\bibinfo {year}
  {2004})%
  \bibAnnoteFile{NoStop}{kostylev04}%
\bibitem{verba12}%
  \BibitemOpen
  \bibfield{author}{%
  \bibinfo {author} {\bibfnamefont{R.}~\bibnamefont{Verba}}, \bibinfo {author}
  {\bibfnamefont{G.}~\bibnamefont{Melkov}}, \bibinfo {author}
  {\bibfnamefont{V.}~\bibnamefont{Tiberkevich}},\ and\ \bibinfo {author}
  {\bibfnamefont{A.}~\bibnamefont{Slavin}},\ }%
  \bibfield{journal}{%
  \Doi{10.1103/PhysRevB.85.014427}{\bibinfo {journal} {Phys. Rev. B}}\ }%
  \textbf{\bibinfo {volume} {85}},\ \bibinfo {pages} {014427} (\bibinfo {year}
  {2012})%
  \bibAnnoteFile{NoStop}{verba12}%
\bibitem{tandon04}%
  \BibitemOpen
  \bibfield{author}{%
  \bibinfo {author} {\bibfnamefont{S.}~\bibnamefont{Tandon}}, \bibinfo {author}
  {\bibfnamefont{M.}~\bibnamefont{Beleggia}}, \bibinfo {author}
  {\bibfnamefont{Y.}~\bibnamefont{Zhu}},\ and\ \bibinfo {author}
  {\bibfnamefont{M.}~\bibnamefont{De~Graef}},\ }%
  \bibfield{journal}{%
  \bibinfo {journal} {J. Magn. Magn. Mater.}\ }%
  \textbf{\bibinfo {volume} {271}},\ \bibinfo {pages} {9} (\bibinfo {year}
  {2004})%
  \bibAnnoteFile{NoStop}{tandon04}%
\bibitem{gurevich96}%
  \BibitemOpen
  \bibfield{author}{%
  \bibinfo {author} {\bibfnamefont{A.~G.}\ \bibnamefont{Gurevich}}\ and\
  \bibinfo {author} {\bibfnamefont{G.~A.}\ \bibnamefont{Melkov}},\ }%
  \emph{\bibinfo {title} {Magnetization Oscillations and Waves}}\ (\bibinfo
  {publisher} {CRC Press},\ \bibinfo {year} {1996})%
  \bibAnnoteFile{NoStop}{gurevich96}%
\bibitem{SpinFlow3D}%
  \BibitemOpen
  \url{http://www.insilicio.fr/pdf/Spinflow_3D.pdf}%
  \bibAnnoteFile{NoStop}{SpinFlow3D}%
\bibitem{sukhostavets11}%
  \BibitemOpen
  \bibfield{author}{%
  \bibinfo {author} {\bibfnamefont{O.~V.}\ \bibnamefont{Sukhostavets}},
  \bibinfo {author} {\bibfnamefont{J.~M.}\ \bibnamefont{Gonzalez}},\ and\
  \bibinfo {author} {\bibfnamefont{K.~Y.}\ \bibnamefont{Guslienko}},\ }%
  \bibfield{journal}{%
  \Doi{10.1143/APEX.4.065003}{\bibinfo {journal} {Applied Physics Express}}\ }%
  \textbf{\bibinfo {volume} {4}},\ \bibinfo {pages} {065003} (\bibinfo {year}
  {2011})%
  \bibAnnoteFile{NoStop}{sukhostavets11}%
\end{thebibliography}
\end{document}